\documentclass[useAMS,usenatbib]{mn2e}

\usepackage{times,url,graphicx}

\begin{document}

%%%%%%%%%%%%%%%%%%%%%%%%%%%%%%%%%%%%%%%%%%%%%%%%%%%%%%%%%%%%%%%%%%%%%%%%%%%%%
%%%%    Title                                                            %%%%
%%%%%%%%%%%%%%%%%%%%%%%%%%%%%%%%%%%%%%%%%%%%%%%%%%%%%%%%%%%%%%%%%%%%%%%%%%%%%

\title[Light curve of PSR B1957+20]{The light curve of the 
companion to PSR B1957+20}
\author[M.~Reynolds et al.]
{M.T.~Reynolds$^{1}$\thanks{email : m.reynolds@ucc.ie}, 
P.J.~Callanan$^{1}$, A.S.~Fruchter$^{2}$, M.A.P.~Torres$^{3}$
\newauthor M.E.~Beer$^{4}$ and R.A.~Gibbons$^{5}$\\
 $^{1}$Physics Department, University College Cork, Ireland\\
 $^{2}$Space Telescope Science Institute, 3700 San Martin Drive, Baltimore, MD 
21218, USA\\
 $^{3}$Harvard-Smithsonian Center for Astrophysics, 60 Garden Street,
Cambridge, MA 02138, USA\\
 $^{4}$Department of Physics and Astronomy, University of Leicester, Leicester 
LE1 7RH, England\\
 $^{5}$Department of Physics and Astronomy, Vanderbilt University,
P.O. Box 1807, Nashville, TN 37240, USA}
                                                                          
\maketitle

%%%%%%%%%%%%%%%%%%%%%%%%%%%%%%%%%%%%%%%%%%%%%%%%%%%%%%%%%%%%%%%%%%%%%%%%%%%%%
%%%%    Abstract                                                         %%%%
%%%%%%%%%%%%%%%%%%%%%%%%%%%%%%%%%%%%%%%%%%%%%%%%%%%%%%%%%%%%%%%%%%%%%%%%%%%%%

\begin{abstract}
We present a new analysis of the light curve for the secondary star in the 
eclipsing binary millisecond pulsar system PSR B1957+20. Combining previous
data and new data points at 
minimum from the Hubble Space Telescope, we have 100\% coverage in the
R-band. We also have a number of new K$_s$-band data points, which we
use to constrain the infrared magnitude of the system. We model this with
the Eclipsing Light Curve code (ELC). From the modelling with the ELC code we
obtain colour 
information about the secondary at minimum light in BVRI and K. For our best
fit model we are able to constrain the system inclination to
65$^{\degr} \pm$ 2$^{\degr}$ for pulsar masses ranging from 1.3 --
1.9 M$_{\sun}$. The pulsar mass is unconstrained. We also find that
the secondary star is not filling its Roche lobe. The temperature of the
un-irradiated side of the companion is in agreement with previous
estimates and we find that the observed temperature gradient across the
secondary star is physically sustainable.
\end{abstract} 
                         

%%%%%%%%%%%%%%%%%%%%%%%%%%%%%%%%%%%%%%%%%%%%%%%%%%%%%%%%%%%%%%%%%%%%%%%%%%%%%
%%%%    Introduction                                                     %%%%
%%%%%%%%%%%%%%%%%%%%%%%%%%%%%%%%%%%%%%%%%%%%%%%%%%%%%%%%%%%%%%%%%%%%%%%%%%%%%
\section{Introduction}
The binary millisecond pulsar (MSP) PSR B1957+20~\citep*{b9} is the 
original and one of the best studied members of its class. It consists
of a 1.6 ms radio pulsar orbiting a companion of mass no
less than $0.022M_{\sun}$, in a binary of orbital period 9.17
hours. For 10\% of this orbit, the radio emission from the pulsar is
eclipsed: the eclipsing region is considerably larger than the Roche
lobe of the companion star, suggesting a wind of material from the
secondary star, due to ablation by the impinging pulsar radiation
~\citep{b9}. The optical counterpart to PSR B1957+20 was discovered by
\citet*{b45};
subsequent observations by \citet*{b2} found the optical counterpart
to vary by a factor of 30-40 in flux over the course of the orbital
period. These observations are hampered by the close proximity of a
line of sight ``contaminator'', only $\sim$ 0.7$\arcsec$ away. 
 
At the time of its discovery it was assumed that PSR B1957+20 was the 
missing link between low-mass X-ray binaries (LMXBs) and isolated millisecond 
pulsars. It was suggested that the high energy pulsar radiation could
evaporate the companion star (Ruderman et al. 1989a \& 1989b) leaving
behind an isolated millisecond pulsar like PSR B1937+21
~\citep{b20}. However, the means by which such a scenario could occur
is still the subject of debate (\citealt*{b23}; see also \citealt{b19},
\citealt{b18} and \citealt{b12} for thorough reviews of the formation and
evolution of MSPs, binary MSPs and the current status of MSP research
respectively). 

PSR B1957+20 is a member of a class of binary pulsar
systems, the Black Widow Pulsars  
(e.g. \citealt{b46}). These are systems with secondaries of mass typically
less than 0.05M$_{\sun}$ and orbital period less than 10 hours. To date
15 such systems have been
identified\footnote{\url{http://www.atnf.csiro.au/research/pulsar/psrcat/exper
    t.html}, \citet{b60}}, with 13 of these residing in globular clusters
and only 2 situated in the field. Radio eclipses have been detected in
approximately half of the cluster systems and in both of the field
systems. Optical light curves exist for only the field systems, PSR
B1957+20 ~\citep{b2} and  PSR J2051-0827 ~\citep{b32,b53}.

Previous attempts at modelling the optical light curve of PSR B1957+20
had one major limitation: the companion was undetectable at
minimum. As such it was impossible to tightly constrain important
system parameters such as the inclination or the degree 
of Roche lobe filling. However, a number of R \& I-band images of
the optical counterpart at minimum were subsequently obtained by the
Hubble Space Telescope (HST). Furthermore, we have recently acquired a number
of K$_s$-band images of the near infrared counterpart. In this paper we
combine these data for the first time to model the light curve to obtain tight
constraints on the inclination and Roche lobe filling fraction.

%%%%%%%%%%%%%%%%%%%%%%%%%%%%%%%%%%%%%%%%%%%%%%%%%%%%%%%%%%%%%%%%%%%%%%%%%%%%%
%%%%    Data                                                             %%%%
%%%%%%%%%%%%%%%%%%%%%%%%%%%%%%%%%%%%%%%%%%%%%%%%%%%%%%%%%%%%%%%%%%%%%%%%%%%%%

\section{Data}

\subsection{Optical Photometry}The data set consists of B, V and R-band
images taken with the William Herschel Telescope (WHT) at La Palma on
the nights of 1989 July 2-5 (see~\citealt{b2} and references therein for
details of the observations), along with two pairs of R (1994 Aug 30,
Oct 28) \& I-band (1994 Sep 03, Oct 27) data points taken at minimum
with the HST. 

We re-reduced the WHT data using the \textsc{IRAF}\footnote{IRAF is distributed
  by the National Optical Astronomy Observatories, which are operated
  by the Association of Universities for Research in Astronomy Inc.,
  under cooperative agreement with the National Science Foundation.}
implementation of \textsc{daophot} \citep{b63}. We undertook this task
as, in their original analysis, ~\citet{b2} used only approximately half
of their data due to poor seeing during 2 of the 4 nights of their
observing run. Our re-analysis  allows us to use an additional night of
data. In the end we had 41 useful R-band data-points covering $\sim$
85\% of the orbit, 39 V-band data points covering $\sim$ 70\% of orbital
phase and 7 useful B-band points taken near maximum.

Two R-band \& 2 I-band data points were also obtained during eclipse
with the Wide Field Planetary Camera 2 using the F675W and F814W filters
respectively. The exposure time was 600s in each case. These images were
corrected for cosmic ray hits and the object magnitudes were calculated
using the \textsc{qphot} task in \textsc{IRAF}. The F675W and F814W
magnitudes were converted to R and I-band magnitudes in the Johnson
system following the prescription given by \citet{b61}.

The WHT and HST data was then phased according to the radio ephemerides of
\citet{b27} and \citet{b66} respectively. When combined, these points
provide the first complete optical light curve of PSR B1957+20, in
addition to colour information throughout the orbital cycle.                
                                                  
%%%%%%%%%%%%%%%%%%%%%%%%%%%%%%%%%%%%%%%%%%%%%%%%%%%%%%%%%%%%%%%%%%%%%%%%%%%%%

\subsection{IR photometry} 

\begin{figure}
\includegraphics[height=90mm,width=84mm]{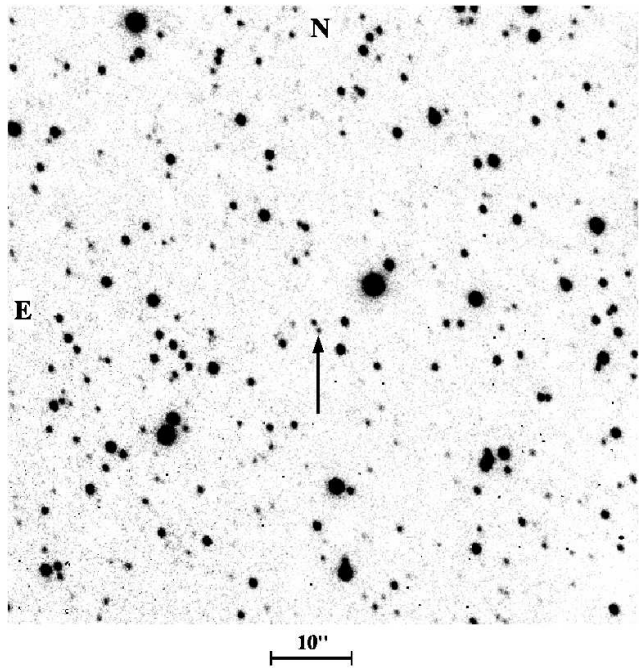} 
\caption{K$_s$ band image of the PSR B1957+20 system. The arrow
  indicates the position of the secondary with the line of sight
  contaminator (0.7$\arcsec$ separation) lying to the north east. The
  exposure time was 180s.}
\label{k-image}
\end{figure}

Our observations consist of a series of K$_s$-band images obtained with
the 6.5m Magellan Baade telescope at Las Campanas Observatory on 2004
July 23 and 2005 September 13 using the PANIC camera. These were
dark-current subtracted, flat fielded, background subtracted and
combined using standard \textsc{IRAF} routines. In total we obtained 4
images totalling approximately 18 minutes on source exposure time. We
display one of our K$_s$-band (1.99$\mu m \ \leq \lambda \leq 2.30 \mu m$
) images in Fig. \ref{k-image}. The companion star to the pulsar is
easily resolved.  

Photometry was carried out in the same manner as the WHT optical
data. The image was calibrated using the standard star
P565-C \citep{b62} and cross checked by comparing a number of stars in
our field with those in the 2MASS catalogue. The data was 
phased using the ephemeris of \citet{b66}, which at the time of our
observations was accurate to at least 1 second. Our final K$_s$-band
photometry is displayed in Table \ref{k-table}.  
 
\begin{table}
\caption{K$_s$ magnitudes of PSR B1957+20.}
\label{k-table}
\begin{center}
\begin{tabular}{ccc}
\hline
$\phi$ &  Secondary K$_s$ mag &  Contaminator K$_s$ mag \\
\hline\hline 
0.5    &  17.8  $\pm$ 0.1   &  18.06 $\pm$ 0.06  \\
0.764  &  18.20 $\pm$ 0.05  &  17.98 $\pm$ 0.04 \\
0.774  &  18.27 $\pm$ 0.10  &  18.13 $\pm$ 0.08 \\
0.788  &  18.35 $\pm$ 0.06  &  18.02 $\pm$ 0.05 \\
0.824  &  18.73 $\pm$ 0.09  &  18.11 $\pm$ 0.07 \\
\hline
\end{tabular}
\end{center}
\end{table}

Previous attempts at IR photometry of this system were unable to
resolve the pulsar from the line of sight contaminator and as such
only the K-band magnitude of the unresolved combination was obtained
(see \citealt{b28}). However, our new observations allow us to subtract the
magnitude of the contaminator from the combined magnitude of
\citet{b28}, yielding a K$_s$-band magnitude of the system at maximum
of 17.8 $\pm$ 0.1.

%%%%%%%%%%%%%%%%%%%%%%%%%%%%%%%%%%%%%%%%%%%%%%%%%%%%%%%%%%%%%%%%%%%%%%%%%%%%
%%%%    The ELC Model                                                   %%%%
%%%%%%%%%%%%%%%%%%%%%%%%%%%%%%%%%%%%%%%%%%%%%%%%%%%%%%%%%%%%%%%%%%%%%%%%%%%%

\section{The ELC Model}
To model these data we used the ELC light curve modelling code
\citep{b3}. The ELC code is ideally suited for this type of
system as it incorporates the  NEXTGEN low temperature model atmosphere
tables, which are critical for systems like PSR
B1957+20, with a companion of likely temperature $\sim$ 3000 K
\citep{b8}. The ELC code also allows one to fit light  curves on a one
by one or simultaneous basis. In our case this allowed the fitting of
the BVRI and K-band light curves simultaneously.

%%%%%%%%%%%%%%%%%%%%%%%%%%%%%%%%%%%%%%%%%%%%%%%%%%%%%%%%%%%%%%%%%%%%%%%%%%%%%

\subsection{The Model}

\begin{table}
\caption{Orbital Parameters of PSR B1957+20 system used in the ELC 
modelling.}
\label{param-table}
\begin{center}
\begin{tabular}{cc}
\hline
Parameter                       &  Value \\
\hline\hline 
Companion Mass$^1$              &  0.022 M$_{\sun}$ \\
Orbital Period$^1$              &  9.17 hr \\
log L$_{spindown}$$^1$          &  35.20 erg s$^{-1}$\\
Companion Effective Temp$^2$    &  2800 K \\
Inclination$^3$                 &  50 -- 80$^{\degr}$\\
$f^3$                           &  $\sim$ full\\
\hline
\end{tabular}
\end{center}
\medskip
{$^1$ Fruchter et al.1988 ; $^2$ Fruchter et al. 1995\\ $^3$ Callanan et al. 
1995}
\end{table}

\begin{figure}                                                     
\includegraphics[width=84mm]{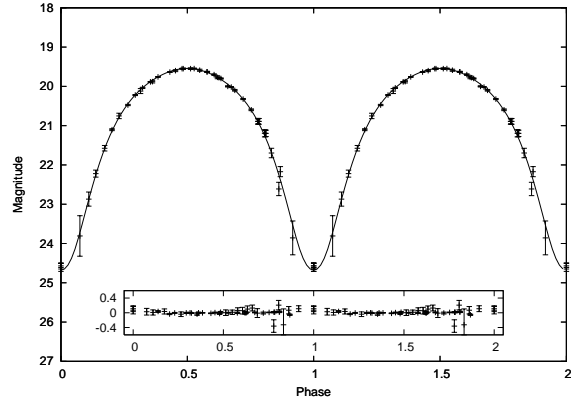} 
\caption{The best fit to the combined R-band data with residuals
  (inset). Two orbital phases are displayed for added clarity. The
  pulsar mass is 1.4 M$_{\sun}$. The best fit inclination is
  \textit{i} = 64.4$^{+1.3\degr} _{-1.2}$ (3$\sigma$) with a $\chi ^2
  _{\nu}$ = 1.06}.  

\label{R-fit}
\end{figure}

\begin{figure}
\includegraphics[width=84mm]{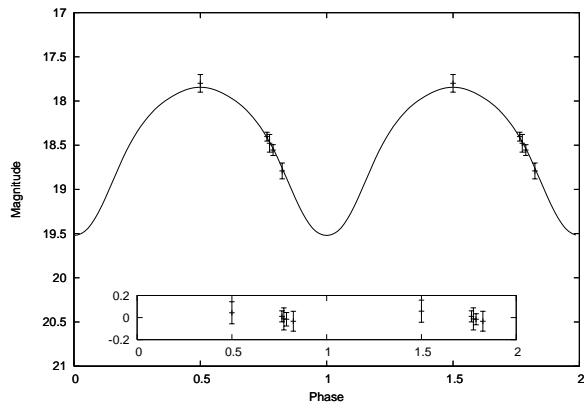} 
\caption{The simultaneous fit to the K-band light curve corresponding to
  the R-band fit in Figure \ref{R-fit}.}
\label{K-fit} 
\end{figure}

The ELC program requires a number of input parameters before modelling
the light curve: the initial parameters used are given in Table
\ref{param-table}. The temperature dependent gravity darkening exponents
of ~\citet{b5} were used. We initially attempted to model the system as
a blackbody (T $\sim$ 2800 K); however, the results were
unsatisfactory. While the code had no problem in fitting the observed
light curve at maximum, it was completely incapable of reproducing the
observed minimum (the model was consistently too luminous during
eclipse). We then employed the NEXTGEN model atmospheres of ~\citet*{b6}
\&~\citet{b7}, and using our blackbody model as our starting point, we
proceeded to model the light curve. The pulsar mass was set to the
canonical value of 1.4 M$_{\sun}$. We then varied the following
parameters: inclination and mass ratio of the system, Roche lobe filling
fraction ($f$), temperature and bolometric albedo ($a$) of the secondary
star and the irradiating luminosity. The
geneticELC algorithm (based on the PIKAIA routine of \citealt{b51}) 
was used to search for the best fit values. The best fit R-band model
is displayed in Figure \ref{R-fit}. We see that there is excellent
agreement between the fit and the data ($\chi ^2_{\nu}$ = 1.06). The
largest deviations occur at orbital phases $\phi > 0.65$, but even these
are well within the errors. This discrepancy is due to the relative
faintness of the companion at these phases, and poorer seeing conditions
during these observations. As a check on the validity of the model we used our
limited K$_s$-band data, as displayed in Figure \ref{K-fit}. We see that
the fit agrees with these data very well. 

Given that the mass of the pulsar is currently unknown, although most
evolutionary scenarios suggest that it will have accreted a few
1/10$^{th}$ of a solar mass from the secondary, we decided to
repeat the above procedure for a number of other primary masses in the
range 1.3 $<$ M$_{MSP}$ $<$ 1.9 M$_{\sun}$, to investigate the
effect of the pulsar's mass on our estimates of the mass ratio and
orbital inclination of the system. 

%%%%%%%%%%%%%%%%%%%%%%%%%%%%%%%%%%%%%%%%%%%%%%%%%%%%%%%%%%%%%%%%%%%%%%%%%%%%%
%%%%      Results                                                        %%%%
%%%%%%%%%%%%%%%%%%%%%%%%%%%%%%%%%%%%%%%%%%%%%%%%%%%%%%%%%%%%%%%%%%%%%%%%%%%%%
\section{Results}

\subsection{Inclination}
\begin{figure}
\includegraphics[height=84mm,angle=-90]{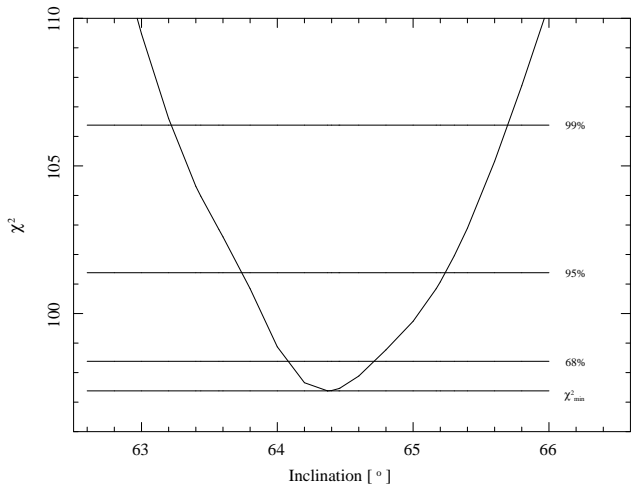}                    
\caption{The graph of $\chi ^2$ vs \textit{i} for a pulsar of mass
1.4 M$_{\sun}$. Minimum  occurs for $\chi ^2 _{\nu}$ = 1.06 and an
inclination, \textit{i} $\sim$ 64.4 $\degr$. The 68\%, 95\% \& 99\%
confidence levels are illustrated.} 
\label{qi1-40}
\end{figure}

At a given pulsar mass the inclination was
constrained to within $\sim \pm$ 1.2$^{\degr}$, i.e. for a pulsar of mass 1.4
M$_{\sun}$, \textit{i} = 64.4$^{+1.3\degr} _{-1.2}$ (see Figure
\ref{qi1-40}), and overall for the above range of pulsar masses we
find the inclination of the system to be in the range, $63^{\degr} \leq i \leq
67^{\degr}$, at the 3$\sigma$ level.
 
%%%%%%%%%%%%%%%%%%%%%%%%%%%%%%%%%%%%%%%%%%%%%%%%%%%%%%%%%%%%%%%%%%%%%%%%%%%%%

\subsection{Pulsar Mass}
The value of $\chi ^2 _{\nu}$ exhibited only a nominal increase as the
mass of the pulsar was increased from 1.3 -- 1.9 M$_{\sun}$:
hence our models are unable to constrain this parameter.

%%%%%%%%%%%%%%%%%%%%%%%%%%%%%%%%%%%%%%%%%%%%%%%%%%%%%%%%%%%%%%%%%%%%%%%%%%%%%

\subsection{Roche lobe filling factor}
At no point in our attempts to model this system were we able to obtain an
acceptable fit for a secondary filling its Roche lobe. For our models
using the NEXTGEN model atmospheres the value of \textit{f} was
approximately constant, 0.81 $\lid$~\textit{f} $\lid$ 0.87 (3$\sigma$
level), as we varied the mass of the pulsar between 1.3 and 1.9
M$_{\sun}$. Hence the secondary is tightly constrained as not
currently filling its Roche lobe.

%%%%%%%%%%%%%%%%%%%%%%%%%%%%%%%%%%%%%%%%%%%%%%%%%%%%%%%%%%%%%%%%%%%%%%%%%%%%%

\subsection{Temperature of the Secondary at maximum and minimum}
We obtained a value of \textit{T} = 2900 $\pm$ 110 K (3$\sigma$ level), 
for the effective temperature of the un-illuminated side of the companion star 
for pulsar masses in the range 1.3 -- 1.9 M$_{\sun}$. For individual pulsar 
masses the 3$\sigma$ error was $\pm$ 90 K i.e. for a pulsar of mass 1.4 
M$_{\sun}$ an effective temperature of \textit{T} = 2900 $\pm$ 90 K was 
obtained. The corresponding temperature at maximum is \textit{T} =
8300 $\pm$ 200 K (3$\sigma$).               
                                                            
From our modelling, we have obtained the magnitude of the secondary at
maximum (in I) and during eclipse (in B, V and K); see Table
\ref{result-table}. These provide us with colour information about
both the cool side and the irradiated side of the pulsar's companion. 

We find that at least $\sim$ 70\% of the spin-down energy of the
pulsar is required to produce the observed heating effect and that
this percentage is independent of the mass of the pulsar. The
bolometric albedo of the system was found to remain close to a value
of 0.5 for all our models, which ensured that the secondary was convective.

\begin{table*}
\caption{Optical and IR magnitudes of the companion to PSR
    B1957+20 at maximum and during eclipse.}
\label{result-table}
\begin{center}
\begin{tabular}{lccccc}
\hline
&B&V&R&I&K \\
\hline\hline 
Max & 21.08 $\pm$ 0.05 & 20.16 $\pm$ 0.05 & 19.53 $\pm$ 0.05 &
\textbf{18.79 $\pm$ 0.05} & 17.8 $\pm$ 0.1 \\
Min &\textbf{28.1 $\pm$ 0.1} & \textbf{26.2 $\pm$ 0.1} & 24.6 $\pm$
0.1 & 22.52 $\pm$ 0.05 & \textbf{19.5 $\pm$ 0.1} \\
\hline
\end{tabular}
\end{center}
\medskip
{The values in bold are predicted by the ELC models, all
  other values are measured directly from the photometry.}
\end{table*}

%%%%%%%%%%%%%%%%%%%%%%%%%%%%%%%%%%%%%%%%%%%%%%%%%%%%%%%%%%%%%%%%%%%%%%%%%%%%%
%%%%      Discussion                                                     %%%%
%%%%%%%%%%%%%%%%%%%%%%%%%%%%%%%%%%%%%%%%%%%%%%%%%%%%%%%%%%%%%%%%%%%%%%%%%%%%%

\section{Discussion}
We have modelled the light curve of the PSR B1957+20 system with the
ELC code and we find the system to be accurately modelled by a highly
irradiated secondary. The inclination of the system is measured to be
65$^{\degr}\pm$ 2$^{\degr}$ for a pulsar in the mass range 1.3 -
1.9 M$_{\sun}$. 

The optical/IR lightcurves display no evidence for the presence of a
contribution from the intra-binary shock. If such a shock did
contribute in a non-negligible manner to the optical/IR flux from the
system, we would expect to observe this in the form of an asymmetrical
distortion of the lightcurves, which is not observed. In contrast the
highly symmetrical nature of the lightcurves is striking evidence that
the modulation is the result of the emission from the heated face of the
secondary star. \citet{b67} observed variable non-thermal X-ray emission
attributed to the intra-binary shock in the binary millisecond pulsar
47 Tuc W\footnote{also known as PSR J0024-7204}; however, an
extrapolation of this emission to optical wavelengths demonstrated that
it contributed negligibly here. Recent \textit{XMM-Newton} observations
\citep{b68} have tentatively detected similar emission in the PSR
B1957+20 system, although our observations show it to have an insignificant
contribution in the optical, as observed in 47 Tuc W.

\subsection{The Roche lobe filling factor}
The Roche lobe filling factor is constrained to be greater than 80\%,
for our best fit model. This result is in agreement with past
estimates that required the secondary to be close to filling its Roche
lobe (\citealt{b10,b15,b17,b2}). Previous estimates of the mass loss
rate in this system \citep{b11} required the system to be close to
filling its Roche lobe, as the measured density of the eclipsing
material was too tenuous to account for significant mass loss. They
claimed that if the secondary neared its Roche lobe, material could
easily leave the stellar surface and remain in the orbital plane; this
would explain the low density of the observed material.

\subsection{Temperature and albedo of the secondary star}
The effective temperature of the un-illuminated side of the secondary,
\textit{T} = 2900 $\pm$ 110 K, is in excellent agreement with the
previous estimate of \textit{T} = 2800 $\pm$ 150 K \citep{b8} and is
corroborated by its agreement with the R - I colour 
temperature obtained via the HST ($\sim$ 3000 K, \citealt{b26}). The
temperature derived from the colour information at maximum is
\textit{T$_{max}$} = 8000$ ^{+1000} _{-3000}$ K. This compares with a
temperature of 8300 $\pm$ 200 K from the ELC models.  The error in the
colour temperature at maximum is dominated by the large uncertainty in
the extinction in the direction of PSR B1957+20. 
  
The bolometric albedo of the secondary, defined as the ratio of the
reradiated energy to the irradiance energy \citep{b65}, was found to
favour the convective case (a $\sim$ 0.5 ) as one would expect for a
secondary of such a small size \citep{b52}. We note here the modelling
of the analogous system PSR J2051-0827 \citep{b53}, in which the
percentage of the pulsar's spin-down luminosity, which is re-radiated in
the optical by the secondary, was determined to be in the region of 30\%
- 45\%. This is consistent with the the incident spin-down luminosity
that our models require for PSR B1957+20 ($\geq$ 70\%), given the above
albedo. 

As a check, a number of models were constructed in which the bolometric
albedo was set to 1.0 (radiative secondary). In this case, the best fit
value of the irradiating luminosity is found to be lower as expected
but the associated temperature of the cool side of the secondary is
inconsistent with the observed colours.

\subsection{Temperature Gradient}
It is clear from the light curve of PSR B1957+20 that a large
temperature gradient is required between the heated and cool
hemispheres of the companion star. To test if this is physically
sustainable, we decided to model the heat flow along the surface of
the secondary in more detail.

A two-dimensional model of the irradiation of PSR B1957+20 was
simulated using a modified version of the code described in \citet*{b35}
. The code uses a polytropic equation of state and only
hydro-dynamical effects were initially considered. The irradiation
causes a stress on the stellar surface which drives a sub-sonic
circulation. Once this circulation pattern was found thermodynamic
effects were considered. Matter in the directly illuminated region was
heated and the advection of this matter was followed across the
stellar surface. As the matter flowed it was allowed to cool
radiatively. The resulting temperature distribution was evolved in
time until a steady-state solution was achieved. It was found that the
heated material extended beyond the directly irradiated region but
that not all of the un-illuminated portion of the star was
heated. Consequently a large temperature gradient between the
illuminated and un-illuminated sides was found to exist.

It may not be obvious how
such a large temperature gradient can exist across the surface of the
secondary. In fact the converse has also been argued. If the radiative
cooling timescale is short then little or no energy would be
redistributed \citep{b33}. The reason why a large temperature gradient
can exist is because the star is perturbed from hydrostatic
equilibrium by the irradiation induced circulation as first noted by
\citet*{b34}. The circulation pattern attempts to distribute the
energy due to heating across the surface of the secondary. However,
the circulation itself, produces inertia terms in the equations of
motion which perturb the star from hydrostatic equilibrium i.e. the
pressure gradient is no longer in the same direction as the potential
gradient. This in turn is what allows the large temperature gradient
to exist across the surface, even of a star as small as the secondary
present in this system. This effect is independent of whether the
object in question is degenerate or non-degenerate.

\subsection{Nature of the secondary}
We can use the above colour/temperature information to constrain the
nature of the secondary star. The colour information appears to rule
out the possibility that the secondary is a white dwarf. In their study
of ultracool white dwarfs (T $<$ 4000 K) \citet{b43} found that white
dwarfs at this low temperature typically have R-I $<$ 0.5, whereas we
find (R-I)$_0$ = 1.8 $\pm$ 0.3. The reddening in the direction of PSR
B1957+20 was calculated using the hydrogen column density estimate of
\citet{b29}, N$_H$ = (1.8 $\pm$ 0.7)$ \times 10^{21}$ cm $^{-2}$, in
combination with the extinction curve of \citet{b37}. Furthermore, if
the secondary in PSR B1957+20 was a $\sim$ 0.025 M$_{\sun}$ white dwarf,
one would expect a radius of $\sim$ 0.1 R$_{\sun}$ - again in contrast
with the value of $\sim$ 0.3 R$_{\sun}$ which we have determined from
our modelling. 

According to \citet{b38} R - I is the most reliable
spectral type indicator for late M-type stars and using their
diagrams of both spectral type \& temperature vs R - I for a sample of
late M dwarfs, we find that temperatures of between 2900 - 3100 K and
spectral types of M4 - M7 are in agreement with our observed R - I.
Hence the secondary appears to exhibit the colours of a late M-type
dwarf, although a main sequence companion (M$_{comp} >$ 0.08 M$_{\sun}$) is
ruled out on the basis of the mass function combined with our
inclination estimate above. It is clear that the current mass of the
secondary is well below the hydrogen burning limit of 0.08 M$_{\sun}$. Hence,
the most likely current state of the secondary is that of a brown
dwarf. 

The low mass secondary in this system has been observed to have a
temperature of $\sim$ 2900 K and a radius encompassing $\sim$ 80\% of
its Roche lobe. In contrast a 50 Myr old 0.025 M$_{\sun}$ brown dwarf would be
expected to have a temperature of $\sim$ 2200 K and a radius approximately
half the size of that observed \citep{b56}. \citet{b15} have
previously proposed a model in which the secondary star is heated to
this temperature through tidal heating; this model also has
the advantage of naturally explaining the orbital period
variability observed by \citet{b66}. However, this model requires the
secondary star to be close to filling its Roche lobe and given that we
observe the secondary to be underfilling its Roche lobe by up to 20\%,
it is questionable if tidal heating would be an efficient heating
mechanism in this scenario. 

The accreting millisecond pulsar SAX J1808.4-3658 is also observed to
have a bloated, low mass ($\sim$ 0.05M$_{\sun}$) companion. In this
case \citet{b44} suggest that the secondary star is 'pumped up' to the
bloated higher entropy state by the heating effect of the thermal
radiation emitted by the neutron star in quiescence. One could
envisage the secondary star in PSR B1957+20 being affected in a
similar manner but in this case the heating would be caused by the
incident spin-down radiation from the radio pulsar.
In reality the situation is likely to be a complicated interplay
between both mechanisms which combine to produce the abnormal secondary
present in this system.

\subsection{Comparison with other MSPs}
At this point we should also compare our results with the other
ablating field system PSR J2051-0827 \citep{b32,b53}, which is
remarkably similar to PSR B1957+20. The secondary in this system
appears to be similar to that in PSR B1957+20. It has a mass of $\sim$
0.025 M$_{\sun}$ and the temperature of the cool side has been measured
to be $\sim$ 3000 K. This system has also been found to
be under-filling its Roche lobe, in this case by $\sim$ 50\%. 
Even though the orbital period of this system is only 2.4 hrs (in
comparison to 9.1 hrs for PSR B1957+20), it is clear that a similar
ablation mechanism is at work.

Interest in these ablating systems has increased with the discovery of
the accretion powered X-ray millisecond pulsars (AXMPs). There is
evidence that the pulsar in these systems is in the process of
'turning-on' as a radio pulsar. In the AXMP SAX J1808.4-3658 \citep{b55},
consisting of a neutron star and a secondary of mass $\sim$ 0.05 M$_{\sun}$
\citep{b44}, a similar process could be taking place. Once again the
secondary is being heated, but in this case the cool side of the
secondary is only 1000 K cooler than the warm side. \citet{b36} have
interpreted this as evidence that the companion star is being irradiated
by the spin down luminosity of the pulsar. Further observations of
quiescent AXMPs are required to test this hypothesis (e.g. \citealt{b59}).

%%%%%%%%%%%%%%%%%%%%%%%%%%%%%%%%%%%%%%%%%%%%%%%%%%%%%%%%%%%%%%%%%%%%%%%%%%%%%
%%%%    Conclusions                                                      %%%%
%%%%%%%%%%%%%%%%%%%%%%%%%%%%%%%%%%%%%%%%%%%%%%%%%%%%%%%%%%%%%%%%%%%%%%%%%%%%%

\section{Conclusions}
The main aim of this paper was to constrain the inclination of the
system as a precursor to a campaign of phase resolved spectroscopy,
with the aim of measuring the mass of the pulsar. We have determined
the inclination to within $\pm 2\degr$ for a pulsar mass in the range
1.3 -- 1.9 M$_{\sun}$. This should ensure that any mass determination will
be limited only by the accuracy of the radial velocity measurements.
We have shown that the temperature of the secondary agrees
with previous estimates and the observed temperature gradient is
physically sustainable. We also find the secondary to be under-filling
its Roche lobe by up to 20\%.\\

\bigskip
 
We thank Jerome Orosz for kindly providing us with the ELC code. 

This paper includes data gathered with the 6.5 meter Magellan Telescopes
located at Las Campanas Observatory, Chile. This research has utilized
2MASS data products. The Two Micron All Sky Survey is a joint project of
the University of Massachusetts and the Infrared Processing and Analysis
Centre/California Institute of Technology, funded by NASA and the
National Science Foundation.  This research made extensive use of the
SIMBAD database, operated at CDS, Strasbourg, France and NASA's
Astrophysics Data System. 

MTR and PJC acknowledge financial support from Science Foundation Ireland.
MAPT was supported in part by NASA LTSA grant NAG5-10889.
%%%%%%%%%%%%%%%%%%%%%%%%%%%%%%%%%%%%%%%%%%%%%%%%%%%%%%%%%%%%%%%%%%%%%%%%%%%%%
%%%%    References                                                       %%%%
%%%%%%%%%%%%%%%%%%%%%%%%%%%%%%%%%%%%%%%%%%%%%%%%%%%%%%%%%%%%%%%%%%%%%%%%%%%%%

%%%%%%%%%%%%%%%%%%%%%%%%%%%%%%%%%%%%%%%%%%%%%%%%%%%%%%%%%%%%%%%%%%%%%%%%%%%%%
%%%%%%%%%%%%%%%%%%%%%%%%%%%%%%%%%%%%%%%%%%%%%%%%%%%%%%%%%%%%%%%%%%%%%%%%%%%%%

\bsp

%%%%%%%%%%%%%%%%%%%%%%%%%%%%%%%%%%%%%%%%%%%%%%%%%%%%%%%%%%%%%%%%%%%%%%%%%%%%%
%%%%%%%%%%%%%%%%%%%%%%%%%%%%%%%%%%%%%%%%%%%%%%%%%%%%%%%%%%%%%%%%%%%%%%%%%%%%%


\begin{thebibliography}{}   %last one was 68

\bibitem[\protect\citeauthoryear{Aldcroft, Romani \& Cordes}{Aldcroft
    et al.}{1992}]{b10} Aldcroft T.L., Romani R.W., Cordes J.M.,
    1992, ApJ, 400, 638

\bibitem[\protect\citeauthoryear{Applegate \& Shaham}{Applegate et
    al.}{1994}]{b15} Applegate J.H., Shaham J., 1994, ApJ, 436, 312

\bibitem[\protect\citeauthoryear{Arzoumanian et al.}{1994}]{b66}
  Arzoumanian Z., Fruchter A.S., Taylor J.H., 1994, ApJ, 426, 85

\bibitem[\protect\citeauthoryear{Backer et al.}{1982}]{b20} Backer D,
  Kulkarni S, Heiles C, Davies M, Goss W.M., 1982, Nature, 300, 615

\bibitem[\protect\citeauthoryear{Beer \& Podsiadlowski}{Beer et
    al.}{2002}]{b35} Beer M.E., Podsiadlowski Ph., 2002, MNRAS, 335, 358

\bibitem[\protect\citeauthoryear{Bessell}{1991}]{b38} Bessell M.S.,
  1991, AJ, 101, 662

\bibitem[\protect\citeauthoryear{Bhattacharya \& van den
    Heuvel}{1991}]{b19} Bhattacharya D., van den Heuvel E.P.J., 1991,
    Phys. Rep., 203, 1   

\bibitem[\protect\citeauthoryear{Bildsten \& Chakrabarty}{2001}]{b44}
  Bildsten L., Chakrabarty D., 2001, ApJ, 557, 292

\bibitem[\protect\citeauthoryear{Bogdanov et al.}{2005}]{b67} Bogdanov
  B., Grindlay J., van den Berg M., 2005, ApJ, 630, 1029 

\bibitem[\protect\citeauthoryear{Brookshaw \& Tavani}{Brookshaw et
    al.}{1995}] {b17} Brookshaw L., Tavani M., 1995, in Fruchter A.S.,
    Tavani M. \& Backer D.C., eds, Millisecond pulsars. A decade of
    surprise, ASPConf Ser, San Francisco

\bibitem[\protect\citeauthoryear{Callanan, Charles \& van
    Paradijs}{Callanan et al.}{1989}]{b1} Callanan P.J., Charles P.A,
    van Paradijs J., 1989, MNRAS, 240, 31

\bibitem[\protect\citeauthoryear{Callanan, van Paradijs \&
    Rengelink}{Callanan  et al.}{1995}]{b2} Callanan P.J.,  van
    Paradijs J., Rengelink R., 1995, ApJ,  439, 928

\bibitem[\protect\citeauthoryear{Callanan et al.}{2007}]{b59} Callanan
  P.J., Reynolds M.T., Filippenko A.V., Foley R., Garnavich P.M., 2007,
  in prep.

\bibitem[\protect\citeauthoryear{Campana et al.}{2004}]{b36} Campana
  S., D'Avanzo P., Casares J., Covino S., Israel G., Marconi G., Hynes
  R., Charles P., Stella L., 2004, ApJ, 614, 49

\bibitem[\protect\citeauthoryear{Chabrier et al.}{2000}]{b56} Chabrier
  G., Baraffe I., Allard F., Hauschildt P., 2000, ApJ, 542, 464 

\bibitem[\protect\citeauthoryear{Charbonneau}{1995}]{b51} Charbonneau
  P., 1995, ApJS, 101, 309

\bibitem[\protect\citeauthoryear{Claret}{2000}]{b5} Claret A., 2000,
  A\&A, 359, 289

\bibitem[\protect\citeauthoryear{Cox}{2000}]{b26} Cox A.N., 2000, in Cox
  A.N., ed., Allen's Astrophysical Quantities, 4$^{th}$
  edn. AIP/Springer-Verlag, New York  

\bibitem[\protect\citeauthoryear{Dahab}{1974}]{b33} Dahab R.E., 1974,
  ApJ, 187, 351

\bibitem[\protect\citeauthoryear{Eales, Becklin, Zuckerman \&
    McLean}{Eales et al.}{1990}]{b28} Eales S.A., Becklin E.E.,
    Zuckerman B., McLean I.S., 1990, MNRAS, 242 , 17

\bibitem[\protect\citeauthoryear{Eichler \& Levinson}{1988 \&
    1991}]{b23} Eichler D., Levinson A., 1988, ApJ, 335, L67

\bibitem[\protect\citeauthoryear{Fruchter, Stinebring \&
    Taylor}{Fruchter  et al.}{1988}]{b9} Fruchter A.S., Stinebring
    D.R., Taylor J.H., 1988, Nature, 333, 237

\bibitem[\protect\citeauthoryear{Fruchter \& Goss}{1992}]{b11}
  Fruchter A.S., Goss W.M., 1992, ApJ, 384, L47 

\bibitem[\protect\citeauthoryear{Fruchter, Bookbinder \&
    Bailyn}{Fruchter et  al.}{1995}]{b8} Fruchter A.S., Bookbinder J.,
    Bailyn C.D., 1995, ApJ, 443, 21 

\bibitem[\protect\citeauthoryear{Gates et al.}{2004}]{b43} Gates E.,
  Gyuk G., Harris H.C. et al., 2004, ApJ, 612, 129

\bibitem[\protect\citeauthoryear{Hauschildt, Allard \&
    Baron}{Hauschildt et al. }{1999a}]{b6} Hauschildt, P.H., Allard
    F., Baron E., 1999, ApJ, 512, 377 

\bibitem[\protect\citeauthoryear{Hauschildt et al.}{1999b}]{b7}
  Hauschildt P.H., Allard F., Ferguson J., Baron E., Alexander D.R.,
  1999, ApJ, 525, 871 

\bibitem[\protect\citeauthoryear{Holtzman et al.}{1995}]{b61} Holtzman
  J.A., Burrows C.J., Casertano S., Hester J.J., Trauger J.T., Watson
  A.M., Worthey G., 1995, PASP, 107, 1065

\bibitem[\protect\citeauthoryear{Huang \& Becker}{Huang et
    al.}{2007}]{b68} Huang H.H., Becker W., 2007, A\&A, 463, 5

\bibitem[\protect\citeauthoryear{in't Zand et al.}{1998}]{b55} in't
  Zand J.J.M., Heise J., Muller J.M., Bazzano A., Cocchi M., Natalucci
  L., Ubertini P., 1998, A\&A, 331, 25

\bibitem[\protect\citeauthoryear{King, Davies \& Beer}{2003}]{b46}
  King A.R., Davies M.B., Beer M.E., 2003, MNRAS, 345, 678

\bibitem[\protect\citeauthoryear{King et al.}{2005}]{b52} King A.R.,
  Beer M.E., Rolfe D.J., Schenker K., Skipp J.M., 2005, MNRAS, 358, 1501

\bibitem[\protect\citeauthoryear{Kippenhahn \& Thomas}{Kippenhahn et
    al.}{1979}]{b34} Kippenhahn R., Thomas H.C., 1979, A\&A, 75, 281

\bibitem[\protect\citeauthoryear{Kulkarni, Djorgovski \&
    Fruchter}{1988}]{b45} Kulkarni S.R., Djorgovski S., Fruchter A.S.,
    1988, Nature, 334, 504

\bibitem[\protect\citeauthoryear{Levinson\& Eichler}{1991}]{b24}
  Levinson A., Eichler D., 1991, ApJ, 379, 359

\bibitem[\protect\citeauthoryear{Leggett}{1992}]{b30} Leggett S.K.,
  1992, ApJS, 82, 351 

\bibitem[\protect\citeauthoryear{Lorimer}{2001}]{b12} Lorimer D.R.,
\url{http://www.livingreviews.org/Articles/Volume4/2001-5lorimer/}

\bibitem[\protect\citeauthoryear{Manchester et al.}{2005}]{b60}
  Manchester R. N., Hobbs G. B., Teoh A., Hobbs, M., 2005, AJ, 129, 1993  

\bibitem[\protect\citeauthoryear{Orosz \& Hauschildt}{Orosz et
    al.}{2000}]{b3} Orosz J.A., Hauschildt P.H., 2000, A\&A, 364, 265.

\bibitem[\protect\citeauthoryear{Persson et al.}{1998}]{b62} Persson
  S.E., Murphy D.C, Krzeminski W., Roth M., Rieke M.J., 1998, AJ, 116, 2475

\bibitem[\protect\citeauthoryear{Phinney \& Kulkarni}{1994}]{b18}
  Phinney E.S., Kulkarni S.R., 1994, ARA\&A, 32, 591

\bibitem[\protect\citeauthoryear{Ruderman Shaham \& Tavani}{Ruderman
    et al.} {1989}]{b21} Ruderman M., Shaham J., Tavani M., 1989a,
    ApJ, 336, 507   

\bibitem[\protect\citeauthoryear{Ruderman, Shaham, Tavani \&
 Eichler}{Ruderman et al.}{1989}]{b22} Ruderman M., Shaham J., Tavani
 M., Eichler D., 1989b, ApJ, 343, 292 

\bibitem[\protect\citeauthoryear{Ryba \& Taylor}{1991}]{b27} Ryba M.F., 
 Taylor J.H.,  1991, ApJ, 380, 557

\bibitem[\protect\citeauthoryear{Savage \& Mathis}{Savage et
    al.}{1979}]{b37} Savage B.D., Mathis J.S., ARA\&A, 17, 73

\bibitem[\protect\citeauthoryear{Stappers, van Kerkwijk, Lane \&
    Kulkarni}{Stappers et al.}{1999}] {b32} Stappers B.W., van
    Kerkwijk M.H., Lane B., Kulkarni S.R.,  1999, ApJ, 510, 45L

\bibitem[\protect\citeauthoryear{Stappers et al.}{2001}]{b53} Stappers
  B.W., van Kerkwijk M.H., Bell J.F., Kulkarni S.R., 2001, ApJ, 548, 183

\bibitem[\protect\citeauthoryear{Stappers, Gaensler, Kaspi, van der
    Klis \& Lewin}{Stappers et al.}{2003}]{b29} Stappers B.W.,
    Gaensler B.M., Kaspi V.M., van der Klis M., Lewin W.H.G., 2003,
    Sci, 299, 1372 

\bibitem[\protect\citeauthoryear{Stetson}{1987}]{b63} Stetson P, 1987,
  PASP, 99, 101

\bibitem[\protect\citeauthoryear{Wilson}{1990}]{b65} Wilson R.E., 1990,
  ApJ, 356, 613

\end{thebibliography}
\end{document}